\documentclass[12pt]{article}
\usepackage{cite}
\usepackage{enumerate}
\usepackage{ifpdf}
\usepackage{ae}
\usepackage[T1]{fontenc}
\usepackage[ansinew]{inputenc}
\usepackage{amsmath}
\usepackage{relsize}
\usepackage{amssymb}
\usepackage{tikz}
\usepackage{graphicx}
\usepackage{color}
\definecolor{darkblue}{cmyk}{0.9,0.9,0,0}
\definecolor{darkgreen}{rgb}{0,0.55,0}
\usepackage[colorlinks=true,linkcolor=darkblue,citecolor=darkblue,urlcolor=darkblue]{hyperref}
\usepackage{epsfig}
\usepackage{epstopdf}
\usepackage{graphicx}

\makeatletter
\long\def\@makecaption#1#2{
  \vskip\abovecaptionskip
  \sbox\@tempboxa{{\captionfonts #1: #2}}
  \ifdim \wd\@tempboxa >\hsize
    {\captionfonts #1: #2\par}
  \else
    \hbox to\hsize{\hfil\box\@tempboxa\hfil}
  \fi
  \vskip\belowcaptionskip}
\makeatother

\newcommand{\beq}{\begin{equation}}
\newcommand{\eeq}{\end{equation}}
\newcommand{\beqy} {\begin{eqnarray}}
\newcommand{\eeqy} {\end{eqnarray}}
\newcommand{\bsmat}{\begin{smallmatrix}}
\newcommand{\esmat}{\end{smallmatrix}}
\newcommand{\bmat}{\begin{matrix}}
\newcommand{\emat}{\end{matrix}}

\def\({\left(}
\def\){\right)}
\def\[{\left[}
\def\]{\right]}

\def\<{\langle}
\def\>{\rangle}

\usepackage{varioref}
\usepackage{makeidx}
\makeindex

\usepackage[english]{babel}
\usepackage{caption}
\usepackage{subfig}

\usepackage{tabularx}
\begin{document}

\thispagestyle{empty}

\renewcommand{\thefootnote}{\fnsymbol{footnote}}
\setcounter{page}{1}
\setcounter{footnote}{0}
\setcounter{figure}{0}
\begin{titlepage}

\begin{center}

\vskip 2.3 cm 

\vskip 5mm

{\Large \bf
Solving CFTs with Weakly Broken Higher Spin Symmetry
}
\vskip 0.5cm

\vskip 15mm

\centerline{Luis F. Alday}
\bigskip
\centerline{\it Mathematical Institute, University of Oxford,} 
\centerline{\it Woodstock Road, Oxford, OX2 6GG, UK}

\end{center}

\vskip 2 cm

\begin{abstract}
\noindent The method of large spin perturbation theory allows to analyse conformal field theories (CFT) by turning the crossing equations into an algebraic problem. We apply this method to a generic CFT with weakly broken higher spin (HS) symmetry, to the first non-trivial order in the breaking parameter. We show that the spectrum of broken currents, for any value of the spin, follows from crossing symmetry. After discussing a generic model of a single scalar field, we focus on vector models with $O(N)$ global symmetry. We rediscover the spectrum of several models, including the $O(N)$ Wilson-Fisher model around four dimensions, the large $O(N)$ model in $2<d<4$ and cubic models around six dimensions, not necessarily unitary. We also discuss models where the fundamental field is not part of the spectrum. Examples of this are weakly coupled gauge theories and our method gives an on-shell gauge invariant way to study them. At first order in the coupling constant we show that again the spectrum follows from crossing symmetry, to all values of the spin. Our method provides an alternative to usual perturbation theory without any reference to a Lagrangian.  

\end{abstract}

\end{titlepage}


\setcounter{page}{1}
\renewcommand{\thefootnote}{\arabic{footnote}}
\setcounter{footnote}{0}

 \def\nref#1{{(\ref{#1})}}
 
\section{Introduction}
Over the last few years there has been great interest in understanding the general structure of the spectrum of dimensions in interacting conformal field theories (CFT). An interesting question is which universal features can be understood from general assumptions, such as conformal invariance and associativity of the operator algebra \cite{Ferrara:1973yt,Polyakov:1974gs,Rattazzi:2008pe}. A beautiful result along these lines is the additivity property for the large spin sector of a generic $CFT_{d>2}$  \cite{Komargodski:2012ek,Fitzpatrick:2012yx}. If the theory contains two operators with twists $\tau_1$ and $\tau_2$, then it also contains an infinite tower of  operators whose twist approaches $\tau_1+\tau_2$ for large spin. This is a nice result, but it also implies that the spectrum of a generic interacting CFT is quite complicated. In order to understand this let us consider for simplicity a single scalar field $\varphi$. Associativity of the operator algebra together with the presence of the identity operator in the OPE $\varphi \times \varphi$ implies that this OPE must also contain a tower of higher spin operators 
\begin{equation}
\varphi \times \varphi= 1+ \left[ \varphi,\varphi \right]_{n,\ell} +\cdots
\end{equation}
schematically of the form $\left[ \varphi,\varphi \right]_{n,\ell} \sim \varphi \Box^n \partial_{\mu_1 \cdots \mu_\ell} \varphi$ and whose dimension approaches twice the dimension of $\varphi$ for large values of the spin in a prescribed manner:
\begin{equation}
\Delta_{n,\ell} = 2\Delta_{\varphi} +2n+\ell + \frac{c}{\ell^{2\rho}} + \cdots,
\end{equation}
see also \cite{Alday:2007mf}. However, for the same reason, the CFT should also contain towers of the form $[\varphi,[\varphi,\varphi]_{n',\ell'}]_{n,\ell}$, and so on! In particular it should posses infinite accumulation points in the spectrum of twists.

A method to study CFT around points of large twist degeneracy was introduced in \cite{Alday:2016njk}. The idea is that accumulation points can be studied by introducing twist conformal blocks $H_\tau^{(0)}(u,v) $, which encode the contribution from towers with unbounded spin and a given twist. Schematically, four point correlators admit the following decomposition
\begin{equation}
{\cal G}(u,v) = \sum_{\tau \in acc.} H_\tau^{(0)}(u,v) + \sum_{isolated} u^{\tau/2} g_{\tau,\ell}(u,v)
\end{equation}
$1/\ell$ corrections to accumulation points can be studied by introducing a sequence of functions $H_\tau^{(\rho)}(u,v)$, defined as $H_\tau^{(0)}(u,v)$ with an extra spin dependent insertion $ \sim \ell^{-2\rho}$. By studying the properties of the sequence $H_\tau^{(\rho)}(u,v)$ around $u,v=0$, we can transform the crossing equations into an algebraic problem. Note that this regime can only be reached in Minkowski space-time. This method is most powerful in theories with a small parameter, which further organises the contributions from the accumulation towers mentioned above. 

In this paper we will apply the method of \cite{Alday:2016njk} to study CFT with weakly broken higher spin (HS) symmetry. In the first class of models we assume the existence of a fundamental field satisfying $\partial_\mu \partial^\mu \varphi=0$ and a tower of conserved currents $J^{(\ell)}$ with twist $\tau=d-2$ at the HS symmetric point . As we break HS symmetry by a small parameter $g$, the fundamental field and currents may acquire an anomalous dimension
\begin{eqnarray}
\Delta_{\varphi} &=& \frac{d-2}{2} +g \, \gamma^{(1)}_{\varphi} +\cdots\\
\Delta_{\ell} &=& d-2 + \ell + g \, \gamma^{(1)}_{\ell} + \cdots,
\end{eqnarray}
We study which spectra are consistent with associativity of the operator algebra and the symmetries of the problem. Our method does not rely in a Lagrangian description and is based on general assumptions:
\begin{itemize}
\item HS symmetry at $g=0$, with a fundamental free field and HS conserved currents. 
\item Associativity of the operator algebra.
\item Exact Conservation of the stress tensor.
\item Exact Invariance under any additional global symmetries.
\end{itemize}
Our strategy  is to study the four point correlator of the fundamental field. The solutions to crossing split into two pieces: a solution unbounded in the spin, which can be obtained by following \cite{Alday:2016njk}, plus a solution with finite support in the spin. The later is much simpler to analyse and is particularly simple for the case at hand. As a result, with this simple set of assumptions we are able to rediscover the spectrum, to order $g^2$ and for any values of the spin, of several models studied in the literature, in dimensions $d>2$. 

In the second class of models we assume the fundamental field $\varphi$ is not part of the spectrum. This is the case of weakly coupled gauge theories. We focus on an almost free four-dimensional theory and study the consequences of crossing for the four point correlator of $\varphi^2=J^{(0)}$. Again, crossing symmetry fixes the spectrum of broken currents for all values of the spin. In particular it implies the presence of Harmonic numbers (as opposed to any other functions with logarithmic behaviour) a result observed in several theories where conserved currents are "single-trace" operators. 

This paper is organised as follows. In section \ref{genericmodel} we study a generic model with a single scalar field in $d$ dimensions and no global symmetries. In section \ref{examples} we consider several models with global symmetries and in various dimensions. We are able to rediscover the spectrum of various examples in the literature. Namely, the $O(N)$ Wilson-Fisher model around four dimensions, the large $N$ critical $O(N)$ model in $2<d<4$ and cubic models around six dimensions. In section \ref{gaugetheories} we analyse a model where the fundamental field is not part of the spectrum. The body of the paper includes also some conclusions. Appendix \ref{twistcb} contains a detailed discussion of the relevant twist conformal blocks, while appendix \ref{finitesupport} constructs all relevant solutions to crossing with finite support on the spin.  

\section{CFT with weakly broken HS symmetry}
\label{genericmodel}

\subsection{HS symmetric point}

In a generic $d-$dimensional CFT, the scaling dimensions of primary operators satisfy unitarity bounds. For a scalar operator $\varphi$ and tensors $J^{(\ell)}$ in the symmetric traceless representation of $SO(d)$ these bounds take the form
\begin{eqnarray}
\Delta_{\varphi} &\geq& \frac{d}{2}-1\\
\Delta_\ell &\geq& d-2+\ell,~~~~\ell \ge 1 
\end{eqnarray}
When these inequalities are saturated the operators satisfy a conservation equation. In the case of currents, their divergence vanish $\nabla \cdot J^{(\ell)} =0$ , and hence they are conserved currents. In the case of a scalar operator, it satisfies the equations of motion $\partial_\mu \partial^\mu \varphi=0$. In this paper we will consider a generic $CFT$ with a whole tower of conserved currents, together with a fundamental field $\varphi$ saturating the unitarity bounds. We will use crossing symmetry to compute the spectrum of this theory as we break HS symmetry. Our starting point is the four point function of fundamental fields at the HS symmetric point
\begin{equation}
\langle \varphi(x_1)\varphi(x_2)\varphi(x_3)\varphi(x_4) \rangle = \frac{{\cal G}^{(0)}(u,v)}{x_{12}^{2 \Delta_{\varphi}^{(0)} }x_{34}^{2 \Delta_{\varphi}^{(0)} }}
\end{equation}
where $2\Delta_{\varphi}^{(0)} =d-2$ and we have introduced cross-ratios $u= \frac{x_{12}^2 x_{34}^2}{x_{13}^2x_{24}^2}$ and $v= \frac{x_{14}^2 x_{23}^2}{x_{13}^2x_{24}^2}$. Crossing symmetry reads
\begin{equation}
v^{\Delta_{\varphi}^{(0)}} {\cal G}^{(0)}(u,v) = u^{\Delta_{\varphi}^{(0)}} {\cal G}^{(0)}(v,u) 
\end{equation}
The intermediate states correspond to the identity operator together with the tower of conserved currents of twist $d-2$
\begin{equation}
{\cal G}^{(0)}(u,v) = 1+ \sum_{\ell} a^{(0)}_\ell u^{\frac{d-2}{2}} g_{d-2,\ell}(u,v)
\end{equation}
where $u^{\frac{\tau}{2}} g_{\tau,\ell}(u,v)$ is the corresponding conformal block in $d$ dimensions. Acting with $\partial_\mu \partial^\mu$ on any of the operators inside the correlator leads to a differential equation for ${\cal G}^{(0)}(u,v)$. This differential equation takes a simpler form when expressed in terms of variables $z,\bar z$:
\begin{equation}
\label{Dsat}
{\cal D}_{sat}{\cal G}^{(0)}(z,\bar z)=0,~~~~~
{\cal D}_{sat}=2 z \bar z(\bar z-z) \partial \bar \partial+(d-2)(z^2 \partial - \bar z^2 \bar \partial)
\end{equation}
where $z \bar z =u,(1-z)(1-\bar z)=v$. Note that the operator ${\cal D}_{sat}$ also anihilates each conformal block $u^{\frac{d-2}{2}} g_{d-2,\ell}(u,v)$ separately. The differential equation for the four-point function is solved by
\begin{equation}
\label{HS4pt}
{\cal G}^{(0)}(u,v) = 1+u^{\frac{d-2}{2}}+ \left(\frac{u}{v} \right)^{\frac{d-2}{2}}
\end{equation}
which is also consistent with crossing symmetry. This result can be shown as follows. Note that the singular term as $v \to 0$ arises from the identity operator together with crossing symmetry. Furthermore, given that intermediate operators (apart from the identity operator) have twist $\tau=d-2$, it follows that this is the only singular term. One can then analytically continue in $d$ and consider an expansion for the singular term of the form
\begin{equation}
{\cal G}^{(0)}(u,v)  = \frac{h(u)}{v^{\frac{d-2}{2}}}
\end{equation}
which should be valid to all orders in $v$. Plugging this into the above equation fixes the form of the singular term. The term that does not depend on $v$ can also be fixed by the above equation, up to a constant, which is then fixed by normalising the contribution from the identity operator to $1$. Positive terms in $v$, of the form $v^\alpha h(u)$ are forbidden, since they would leave to singular terms upon $u \to u/v,v\to 1/v$. This leads to the final result (\ref{HS4pt}).This also agrees with the expectation that correlators in a CFT with HS symmetry should coincide with the ones in a free theory \cite{Maldacena:2011jn}. From the explicit answer for the correlator we can compute the OPE coefficients $a_\ell^{(0)}$. In order to do this it is sufficient to know the conformal blocks in the small $u$ limit, given by the collinear conformal blocks in any number of dimensions \cite{Dolan:2011dv}. We find
\begin{equation}
\label{HSope}
a_\ell^{(0)} = (1+(-1)^\ell)\frac{\Gamma \left(\frac{d}{2}+\ell-1\right)^2 \Gamma (d+\ell-3)}{\Gamma \left(\frac{d}{2}-1\right)^2 \Gamma (\ell+1) \Gamma (d+2 \ell-3)},
\end{equation}
one can check that this result agrees with the HS invariant given in \cite{Skvortsov:2015pea}. 

At the HS symmetric point there is a large degeneracy in the twist. In particular, the conserved currents form a tower of intermediate operators, which is unbounded in the spin and has degenerate twist $\tau=d-2$. In such situations it is convenient to define a twist conformal block $H_{\tau}^{(0)}(u,v)$, which resums the contribution to the four-point correlator from a given twist. In this language
\begin{equation}
{\cal G}^{(0)}(u,v) =H_{0}^{(0)}(u,v)+H_{d-2}^{(0)}(u,v),
\end{equation}
where $H_{0}^{(0)}(u,v)=1$ and 
\begin{equation}
H_{d-2}^{(0)}(u,v)=\sum_{\ell} a^{(0)}_\ell u^{\frac{d-2}{2}} g_{d-2,\ell}(u,v)=u^{\frac{d-2}{2}}+ \left(\frac{u}{v} \right)^{\frac{d-2}{2}}
\end{equation}
As an aside remark, note that $H_{0}^{(0)}(u,v)$ and $H_{d-2}^{(0)}(u,v)$ also coincide with the HS conformal blocks, see \cite{Alday:2016njk}. 

Before proceeding let us mention that there are interesting examples with an infinite tower of conserved currents $J^{(\ell)}$ but where the fundamental field is not part of the spectrum. An example of this are weakly coupled gauge theories. In this case the simplest correlator to study is that of four $J^{(0)}$. We will return to this case in section 4. 

\subsection{Breaking HS symmetry - order $g$}
Let us now break HS symmetry. We assume we do so by a small parameter $g$. We furthermore assume that at this order no new operators enter in the OPE $\varphi \times \varphi$. This assumption will be relaxed at the next order. As we turn on $g$, the conserved currents, together with the external operator, may acquire an anomalous dimension:
\begin{eqnarray}
\Delta_{\varphi} &=& \frac{d-2}{2} +g \, \gamma^{(1)}_{\varphi} +\cdots\\
\Delta_{\ell} &=& d-2 + \ell + g \, \gamma^{(1)}_{\ell} + \cdots,
\end{eqnarray}
Correspondingly
\begin{eqnarray}
{\cal G}(u,v) = {\cal G}^{(0)}(u,v) + g \,{\cal G}^{(1)}(u,v) + \cdots
\end{eqnarray}
The crossing condition $v^{\Delta_\varphi} {\cal G}(u,v)=u^{\Delta_\varphi} {\cal G}(v,u) $ at order $g$ reads
\begin{eqnarray}
v^{\frac{d-2}{2}}\left(\gamma^{(1)}_{\varphi} \log v \, {\cal G}^{(0)}(u,v)+ {\cal G}^{(1)}(u,v) \right) = u^{\frac{d-2}{2}}\left(\gamma^{(1)}_{\varphi} \log u \, {\cal G}^{(0)}(v,u)+ {\cal G}^{(1)}(v,u) \right)
\end{eqnarray}
Which can also be written as
\begin{eqnarray}
\label{crossingtoy}
\gamma^{(1)}_{\varphi} \log v \, {\cal G}^{(0)}(u,v)+{\cal G}^{(1)}(u,v) = \gamma^{(1)}_{\varphi} \log u \, {\cal G}^{(0)}(u,v) + \frac{u^{\frac{d-2}{2}}}{v^{\frac{d-2}{2}}} {\cal G}^{(1)}(v,u) 
\end{eqnarray}
Let us analyse corrections to the spectrum consistent with this condition. It is then convenient to extend the twist conformal blocks to the sequence of functions $H_{d-2}^{(\rho)}(u,v)$, defined exactly in the same way, with an extra spin-dependent insertion:
\begin{equation}
H_{d-2}^{(\rho)}(u,v)=\sum_{\ell} J^{-2\rho}a^{(0)}_\ell u^{\frac{d-2}{2}} g_{d-2,\ell}(u,v),~~~~J^2=(\ell+\tau/2)(\ell+\tau/2-1)
\end{equation}
with $\tau=d-2$. The properties of this sequence of functions are studied in appendix \ref{twistcb}. Let us now assume that the anomalous dimensions $\gamma^{(1)}_{\ell}$ admit an expansion in inverse powers of the spin\footnote{This actually follows from crossing symmetry for the present case. In general, this expansion may also contain terms that behave logarithmically with the spin, as will be the case in section \ref{gaugetheories}.}
\begin{equation}
\gamma^{(1)}_{\ell} = 2\sum_\rho \frac{B_\rho}{J^{2\rho}},~~~~J^2=(\ell+d/2-1)(\ell+d/2-2)
\end{equation}
The piece proportional to $\log u$ on the l.h.s. of the crossing relation (\ref{crossingtoy}) then admits the following decomposition
 \begin{equation}
\left. {\cal G}^{(1)}(u,v)\right|_{\log u} = \sum_\rho B_{\rho} H^{(\rho)}_{d-2}(u,v) + \sum_{\ell}^L  a_{\ell}^{(0)} \frac{\gamma_{\ell}^{f.s.}}{2} u^{\frac{d-2}{2}} g_{d-2,\ell}(u,v)
\end{equation}
This contains two contributions. One arising from solutions which are not truncated in the spin, which admit an expansion in $1/\ell$. Another corresponding to solutions $\gamma_{\ell}^{f.s.}$ with finite support on the spin. The spectrum of anomalous dimensions to all orders in $1/\ell$ is encoded in the coefficients $B_{\rho}$. The twist conformal blocks are studied in appendix $A$ where the expansion of $H^{(\rho)}_{d-2}(u,v)$ around $v=0$ is given. Comparing divergences as $v \to 0$ on both sides of (\ref{crossingtoy}), we see that the only solution with unbounded support in the spin corresponds to the uniform solution
\begin{equation}
\gamma_\ell^{(1)} = 2 \gamma_{\varphi}^{(1)}
\end{equation}
This should be valid to all orders in $1/\ell$. Let us now turn to solutions with finite support on the spin. These are studied in appendix \ref{finitesupport}. For $d \neq 4$ there are no solutions. If we furthermore require that the stress-tensor is conserved, namely $\gamma_2^{(1)}=0$, this leads to no non-trivial solutions for $d \neq 4$ at this order. For $d=4$ there is one truncated solution with $\gamma_0^{(1)}= \alpha$, where $\alpha$ is a free parameter. Note that $\gamma_2^{(1)}$ is still unambiguous and hence conservation of the stress-tensor leads to the same constraints as before. In summary, at order $g$ we find
\begin{equation}
\gamma_{\varphi}^{(1)}=0,~~~\gamma^{(1)}_{\ell}=0,~~~\text{for $\ell>0$},~~~~~~\gamma_0^{(1)} =\left\{
\begin{array}{c l}      
     0 & ~~~d \neq 4 \\
   \alpha & ~~~ d=4
\end{array}\right.
\end{equation}

\subsection{Order $g^2$}
Let us now analyse the crossing equations to order $g^2$
\begin{eqnarray}
v^{\frac{d-2}{2}}(1+g^2 \gamma^{(2)}_{\varphi} \log v + \cdots)\left({\cal G}^{(0)}(u,v)+g {\cal G}^{(1)}(u,v)+g^2 {\cal G}^{(2)}(u,v)+ \cdots\right) = \\
u^{\frac{d-2}{2}} (1+g^2 \gamma^{(2)}_{\varphi} \log u + \cdots) \left({\cal G}^{(0)}(v,u)+g {\cal G}^{(1)}(v,u)+g^2 {\cal G}^{(2)}(v,u)+ \cdots\right) \nonumber
\end{eqnarray}
where we have used $\gamma_{\varphi}^{(1)}=0$, determined above. As we have seen, there is a distinction between $d=4$ and $d \neq 4$. Let us analyse first the case $d=4$. From the decomposition in conformal blocks for ${\cal G}^{(2)}(u,v)$ we can read off  the term proportional to $\log^2 u$
 \begin{equation}
{\cal G}^{(2)}(u,v) = \frac{1}{8}\alpha^2 \log^2 u a_{2,0}^{(0)} u g_{2,0}(u,v) + \cdots
\end{equation}
Note that this piece depends only on the solution at order $g$. This leads to the following piece proportional to $\log^2u\log v$
 \begin{equation}
{\cal G}^{(2)}(u,v) = \frac{u}{z-\bar z} \frac{1}{8}\alpha^2 a_{2,0}^{(0)}  \log^2u\log v + \cdots
\end{equation}
where we have used the explicit form of the conformal block $g_{2,0}(u,v)$ in four dimensions. This term leads to the following contribution on the r.h.s of the crossing relation:
\begin{eqnarray}
(1+g^2 \gamma^{(2)}_{\varphi} \log v + \cdots)\left({\cal G}^{(0)}(u,v)+g {\cal G}^{(1)}(u,v)+g^2 {\cal G}^{(2)}(u,v)+ \cdots\right) = \\
\frac{u}{v} (1+g^2 \gamma^{(2)}_{\varphi} \log u + \cdots) \left(1+ g^2 \frac{v}{z-\bar z} \frac{1}{8}\alpha^2 a_{2,0}^{(0)} \log^2 v \log u+ \cdots   \right) \nonumber
\end{eqnarray}
On the r.h.s. there are two terms, proportional to $\log u/v$ and $\log^2 v \log u$ whose divergence as $v \to 0$ is enhanced with respect to the divergence of a single conformal block. Hence, they should arise from an infinite sum over the spins on the l.h.s. In other words, the correction to order $g^2$ to the anomalous dimensions of the currents should be such that
 \begin{equation}
 \sum_\rho B_{\rho} H^{(\rho)}_2(z,\bar z) = \gamma^{(2)}_\varphi\frac{u}{v} - \frac{1}{8}\frac{z \bar z}{\bar z-z} \alpha^2 a_{2,0}^{(0)} \log^2 (1-\bar z)
\end{equation}
Using the results in appendix \ref{twistcb} for the twist conformal blocks in four dimensions, this problem is very easy to solve. We find
\begin{equation}
B_{0} =\gamma^{(2)}_\varphi,~~~B_{1} = \frac{\alpha^2}{4} a_{2,0}^{(0)} =- \frac{\alpha^2}{2},
\end{equation}
while all other $B_\rho$ vanish. This leads to the following anomalous dimensions for the currents at order $g^2$:
\begin{equation}
\gamma_\ell^{(2)} = 2 \gamma^{(2)}_\varphi - \frac{\alpha^2}{\ell(\ell+1)},
\end{equation}
where we have used the form of the collinear spin in this case $J^2 =\ell (\ell+1)$. This result is valid to all orders in $1/\ell$. Actually, from the form of the solutions with finite support, found in appendix \ref{finitesupport}, it follows this result should be valid for all $\ell>0$ and the only freedom is in $\gamma^{(2)}_0=\gamma^{(2)}_{\varphi^2}$. Since the theory is assumed to be conformal, it should have a conserved current of spin two, the stress tensor. This implies $\gamma_2^{(2)}=0$ and provides a further constraint
\begin{equation}
\gamma^{(2)}_\varphi = \frac{\alpha^2}{12}
\end{equation}
In summary, to order $g^2$ we have fixed the spectrum of intermediate operators, and even that of the external operator, in terms of the dimension of $\varphi^2$:
\begin{equation}
\label{spectrum4d}
\gamma_\varphi = \frac{\alpha^2}{12} g^2 + \cdots ~~~,\gamma_\ell = 2 \gamma_\varphi\left(1 - \frac{6}{\ell(\ell+1)} \right) + \cdots,~~~\gamma_{\varphi^2} = \alpha g+ \cdots
\end{equation}
Let us now analyse the case $d \neq 4$. In this case ${\cal G}^{(1)}(u,v)=0$ so that 
\begin{eqnarray}
\label{crossinggs}
\gamma^{(2)}_{\varphi} \log v \, {\cal G}^{(0)}(u,v)+{\cal G}^{(2)}(u,v) = \gamma^{(2)}_{\varphi} \log u \, {\cal G}^{(0)}(u,v) + \frac{u^{\frac{d-2}{2}}}{v^{\frac{d-2}{2}}} {\cal G}^{(2)}(v,u) 
\end{eqnarray}
In order to have non-trivial solutions new operators must appear in the OPE of $\varphi$ with itself at this order. Let us assume we have a new operator $\chi$
\begin{equation}
\varphi \times \varphi = 1+\chi+ J_s 
\end{equation}
This will lead to the following term in ${\cal G}^{(2)}(v,u)$
\begin{equation}
{\cal G}^{(2)}(v,u)  = a_{\varphi^2\chi} v^{\frac{\tau_\chi}{2}} g_{\tau_\chi,\ell_\chi}(v,u) + \cdots
\end{equation}
where the OPE coefficient $a_{\varphi^2\chi}$ is of order $g^2$ by assumption. The presence of this term will generally lead to a divergence on the r.h.s. of (\ref{crossinggs}) which can only be matched with the correct behaviour for $\gamma_\ell^{(2)}$. Note that this is true for generic, not necessarily small $\tau_\chi$: also an irrational power of $v$ can only be obtained by summing an infinite number of conformal blocks. In order to exploit this fact we can look for the term proportional to $\log u$ on both sides of the crossing equation (\ref{crossinggs})
\begin{equation}
\sum B_\rho H_{d-2}^{(\rho)}(u,v)  = \gamma^{(2)}_{\varphi} \, {\cal G}^{(0)}(u,v)+   \left. a_{\varphi^2\chi} u^{\frac{d-2}{2}} \frac{v^\frac{\tau_\chi}{2}}{v^{\frac{d-2}{2}}} g_{\tau_\chi,\ell_\chi}(v,u) \right|_{\log u}
\end{equation}
The first term on the r.h.s will lead to the uniform contribution proportional to $\gamma^{(2)}_{\varphi} $. Matching divergences on both sides implies the sum over $\rho$ must in addition contain $\rho=\frac{\tau_\chi}{2}+m$, $m=0,1,2,\cdots$. The first coefficient $B_{\frac{\tau_\chi}{2}}$ can be fixed by looking at the leading term for small $v$:
 \begin{equation}
B_{\frac{\Delta_\chi}{2}} \left. H_{d-2}^{(\frac{\tau_\chi}{2})}(u,v) \right|_{v \sim 0}   =  -\frac{\Gamma(\Delta_\chi)}{\Gamma(\frac{\Delta_\chi}{2})^2}  a_{\varphi^2\chi}u^{\frac{d-2}{2}} \frac{v^\frac{\tau_\chi}{2}}{v^{\frac{d-2}{2}}}(1-u)^{\ell_\chi}~_2F_1(\frac{\tau_\chi}{2}+\ell_\chi,\frac{\tau_\chi}{2}+\ell_\chi,1;u)
\end{equation}
The $u$ dependence on both sides only matches provided $\chi$ is a scalar operator with $\Delta_\chi=2$ classically\footnote{Crossing symmetry also allows the exchange of operators whose twist differs from $d-2$ by an even integer. None of the conclusions that follow would be affected by their presence.}. Note that in a Gaussian model with a cubic interaction $\mu \varphi^2 \chi$ this fact would follow from Lorentz and classical scale invariance. Here the same result arises from crossing symmetry. We obtain
\begin{equation}
B_{\frac{\Delta_\chi}{2}}= -\frac{(d-4)^2}{4}a_{\varphi^2\chi}
\end{equation}
where we have used $\Delta_\chi=2$. In order to solve for $B_{\frac{\Delta_\chi}{2}+m}$ for $m>0$ we can use the expression for the conformal block of a scalar operator in $d$ dimensions, given in \cite{Dolan:2000ut}, together with the expansions for $H_{d-2}^{(\rho)}(u,v)$, which can be constructed as explained in appendix \ref{twistcb}. Solving for  $B_{\frac{\Delta_\chi}{2}+m}$, it turns out they all vanish except for the first one! We then arrive to the following result
\begin{equation}
\label{sigmaexchange}
\gamma_\ell^{(2)} = 2 \gamma^{(2)}_\varphi -\frac{(d-4)^2}{2}\frac{a_{\varphi^2\chi}}{(\ell+\frac{d-2}{2})(\ell+\frac{d-2}{2}-1)} ,
\end{equation}
valid for all values of $\ell$. If furthermore we impose the constraint of conservation of the spin two current $\gamma_2^{(2)} =0$ we get a further relation between $\gamma^{(2)}_\varphi$ and $a_{\varphi^2\chi}$. In this case
\begin{equation}
\gamma^{(2)}_\varphi  = \frac{(d-4)^2}{d(d+2)}a_{\varphi^2\chi}
\end{equation}
After this general discussion, let us turn our attention to specific examples with extra symmetries. Each example below will turn out to be a full fledged CFT and interesting on its own right. 

\section{Examples}
\label{examples}
In the following we will consider several specific examples with global symmetries.  All these examples will turn out to be equivalent to Lagrangian theories with a small coupling constant $g$, such that at the point $g=0$ the theory is free and possesses HS symmetry. As we turn on $g$ the fundamental fields and conserved currents acquire anomalous dimensions. These anomalous dimensions have been computed by a variety of methods. Our aim is to reproduce these results solely from the crossing equations consistent with the global symmetry in each case, and the set of assumptions mentioned at the introduction. In particular, we will not make reference to the Lagrange description or the equations of motion. 

\subsection{Wilson-Fisher model}
Let us start with a model of $N$ scalar fields $\varphi^i$ with global $O(N)$ symmetry in $d=4-\epsilon$ dimensions. Let us discuss first the case of a single scalar field and repeat the previous discussion around four dimensions\footnote{To the order we consider in the breaking parameter the relevant conformal blocks, and twist conformal blocks, are still four dimensional.}. The general method above can be directly applied to the four point function
\begin{equation}
\langle \varphi(x_1) \varphi(x_2)\varphi(x_3)\varphi(x_4)\rangle = \frac{{\cal G}(u,v)}{x_{12}^{2\Delta_{\varphi}}x_{34}^{2\Delta_{\varphi}}}
\end{equation}
which at the HS symmetric point is
\begin{equation}
{\cal G}^{(0)}(u,v) = 1+ H_2^{(0)}(u,v),~~~~~H_2^{(0)}(u,v)=\frac{u}{v}+u
\end{equation}
The sequence of functions $H_2^{(m)}(u,v)$ can be constructed directly in $d=4$, as done in appendix \ref{twistcb}. We can then consider corrections to first order in $g$. The piece proportional to $\log u$ admits the following expansion
 \begin{equation}
\left. {\cal G}^{(1)}_1(u,v) \right|_{\log u}= \sum_\rho B_{\rho} H^{(\rho)}_2(u,v) + \sum_{\ell}^L a_{\tau,\ell}^{(0)} \gamma_{\ell}^{f.s} u g_{2,\ell}(u,v)
\end{equation}
where the spectrum of the anomalous dimensions to all orders in $1/\ell$ is encoded in the coefficients $B_{\rho}$. An important feature of the four-dimensional twist conformal blocks $H^{(\rho)}_2(u,v)$ is that they always contains a piece proportional to $\log^2 v$ for $\rho>0$. This piece cannot be possibly reproduced by anything on the dual channel. Hence, it has to be absent. As a result, the anomalous dimensions $\gamma^{(1)}_{\ell}$ should only contain a uniform piece $\gamma^{(1)}_{\ell}=2 \gamma^{(1)}_{\varphi}$, exactly as before. Conservation of the stress-tensor then sets $\gamma^{(1)}_{\ell}=2 \gamma^{(1)}_{\varphi}=0$. Note that anomalous dimensions are measured with respect to $d=4-\epsilon$. This result is valid to all orders in $1/\ell$. Note that the absence of non-truncated solutions can be argued by analytically continuing in the parameter $d$, as we did in the previous section, as well as directly in $d=4$, as we did here. On the other hand, we can add a finite support solution leading to
\begin{equation}
\gamma^{(1)}_{\varphi}=0,~~~\gamma^{(1)}_{0} = \alpha,~~~~\gamma^{(1)}_{\ell} = 0,~~~\text{for $\ell>0$}
\end{equation}
The analysis to order $g^2$ is also exactly as before and we end up with the spectrum (\ref{spectrum4d}).

The generalisation to the $O(N)$ model is relatively straightforward. In this case we consider the four point correlator of the fundamental field $\varphi_i$:
\begin{equation}
\langle \varphi_i(x_1)\varphi_j(x_2)\varphi_k(x_3)\varphi_l(x_4)\rangle = \frac{{\cal G}_{ijkl}(u,v)}{x_{12}^{2\Delta_{\varphi}}x_{34}^{2\Delta_{\varphi}}}
\end{equation}
Crossing symmetry implies
\begin{equation}
v^{\Delta_{\varphi}} {\cal G}_{ijkl}(u,v) =u^{\Delta_{\varphi}} {\cal G}_{kjil}(v,u)
\end{equation}
Intermediate states decompose into representations of the $O(N)$ global symmetry
\begin{eqnarray}
{\cal G}_{ijkl}(u,v) = \delta_{ij}\delta_{kl} {\cal G}_S(u,v) +\left( \frac{\delta_{ik}\delta_{jl}+\delta_{il}\delta_{jk}}{2}-\frac{1}{N}\delta_{ij}\delta_{kl} \right){\cal G}_T(u,v)
+ \frac{\delta_{ik}\delta_{jl}-\delta_{il}\delta_{jk}}{2}{\cal G}_A(u,v)
\end{eqnarray}
where $S,T,A$ denote the singlet, traceless symmetric and anti-symmetric representations of $O(N)$ respectively. The crossing relations decompose accordingly. Written in terms of ${\cal G}(u,v) = v^{\Delta_{\varphi}} f(u,v)$ they take the form
\begin{eqnarray}
\label{crossingON}
f_T(u,v)+f_A(u,v)&=&f_T(v,u)+f_A(v,u) \nonumber\\
f_S(u,v) + \left(1-\frac{1}{N}\right)f_T(u,v) &=&f_S(v,u) + \left(1-\frac{1}{N}\right)f_T(v,u),\\
2 f_S(u,v) - \left(1+\frac{2}{N}\right) f_T(u,v) + f_A(u,v)&=& -2 f_S(v,u) +  \left(1+\frac{2}{N}\right) f_T(v,u)-f_A(v,u). \nonumber
\end{eqnarray}
Furthermore, each component admits a decomposition in conformal blocks over the intermediate states transforming in the corresponding representation. At $g=0$ the four-point function is simply
\begin{equation}
{\cal G}^{(0)}_{ijkl}(u,v) = \delta_{ij}\delta_{kl} + u \delta_{ik}\delta_{jl}  + \frac{u}{v} \delta_{il}\delta_{jk}
\end{equation}
which leads to
\begin{equation}
{\cal G}^{(0)}_S(u,v) =1+ \frac{u}{N}\left( 1+ \frac{1}{v} \right),~~~{\cal G}^{(0)}_T(u,v) =u\left( 1+ \frac{1}{v} \right),~~~{\cal G}^{(0)}_A(u,v) = u\left( 1- \frac{1}{v} \right)
\end{equation}
the intermediate operators include the identity operator and the conserved currents in the singlet, symmetric traceless and anti-symmetric representations $J^{(\ell)}_S,J^{(\ell)}_T,J^{(\ell)}_A$, where the spin is even for the first two representations and odd for the anti-symmetric one. From the explicit expression for the correlator at $g=0$ we can perform the expansion in conformal blocks and compute the corresponding OPE coefficients. For the traceless currents the OPE coefficients take exactly the same form (\ref{HSope}), with $d=4$, while the OPE coefficients  for the singlet currents $J^{(\ell)}_S$ have an additional factor $1/N$.

At $g \neq 0$ the currents acquire anomalous dimensions $\gamma_{S,\ell},\gamma_{T,\ell},\gamma_{A,\ell}$. Conservation of the stress tensor implies $\gamma_{S,2}=0$, while conservation of the global current implies $\gamma_{A,1}=0$. The analysis at order $g$ is very similar to the one performed above. One can see that the only solution with infinite support in the spin is the uniform solution, proportional to $\gamma_{\varphi}$. Then conservation of the stress tensor implies $\gamma_{\varphi}=0$ at this order. Again, we have one solution with finite support, which corresponds to an anomalous dimension for $\gamma_{S,0} \equiv \gamma_{\varphi^2_S}$ and $\gamma_{T,0} \equiv \gamma_{\varphi^2_T}$. The last equation in (\ref{crossingON}) gives a relation between them, so that
\begin{equation}
\label{oneloopON}
\gamma_{\varphi^2_S}^{(1)}= \alpha,~~~\gamma_{\varphi^2_T}^{(1)} = \frac{2}{2+N}\alpha
\end{equation}
For some parameter $\alpha$ which measures the departure from the free theory. In order to consider the problem at order $g^2$ it is convenient to write the crossing equations as
\begin{eqnarray}
\label{crossON}
f_S(u,v) &=& \frac{1}{N}f_S(v,u) + \frac{N^2+N-2}{2N^2} f_T(v,u) +\frac{1-N}{2N} f_A(v,u)\nonumber\\
f_T(u,v) &=&f_S(v,u) + \frac{N-2}{2N} f_T(v,u) +\frac{1}{2} f_A(v,u)\\
f_A(u,v) &=&-f_S(v,u) + \frac{2+N}{2N} f_T(v,u) +\frac{1}{2} f_A(v,u) \nonumber
\end{eqnarray}
The one-loop anomalous dimensions (\ref{oneloopON}) produce the term
\begin{equation}
{\cal G}_{S/T}(v,u) = \frac{1}{8} \frac{v}{z-\bar z} \left( \gamma_{\varphi^2}^{(1)}\right)^2 a^{(0)}_{\varphi^2} \log u \log^2 v + \cdots
\end{equation}
on the r.h.s. of the crossing equations, where $\gamma_{\varphi^2}^{(1)}$ and $a^{(0)}_{\varphi^2}$ denote the corresponding anomalous dimension and OPE coefficient (for the singlet or the symmetric traceless representation). From the tree-level results we find $a^{(0)}_{\varphi^2_T}=2,a^{(0)}_{\varphi^2_S}=2/N$. Matching this term leads to the anomalous dimensions for the currents as an expansion in $1/\ell$, to all orders. Since when adding finite-support solutions the only freedom is in the dimension of $\varphi^2$, these results are valid for all $\ell>0$, exactly as discussed in the previous section. We find the following explicit expressions:
\begin{equation}
\gamma^{(2)}_{S,\ell} = 2 \gamma^{(2)}_\varphi-\frac{3}{N+2} \frac{\alpha^2}{\ell(\ell+1)},~~~\gamma^{(2)}_{T,\ell}=2 \gamma^{(2)}_\varphi-\frac{N+6}{N+2} \frac{\alpha^2}{\ell(\ell+1)},~~~\gamma^{(2)}_{A,\ell}=2 \gamma^{(2)}_\varphi-\frac{1}{N+2} \frac{\alpha^2}{\ell(\ell+1)}
\end{equation}
Requiring the stress tensor to be conserved gives a further relation
\begin{equation}
\gamma^{(2)}_{S,2} = 0 \to \gamma^{(2)}_\varphi=\frac{1}{N+2} \frac{\alpha^2}{4}
\end{equation}
which automatically leads to convervation of the $O(N)$ symmetry $\gamma^{(2)}_{A,1}=0$. In summary, we have completely fixed the spectrum of intermediate operators, and the dimension of the external one, in terms of a single parameter, chosen to be $\gamma_{\varphi^2_S}$. 

What should this be compared to? The Wilson-Fisher model is described by $N$ scalar fields in $d=4-\epsilon$ dimensions with action

\begin{equation}
S = \int d^d x \left[\frac{1}{2} \partial_\mu \varphi^i \partial^\mu \varphi^i+ \frac{g \mu^\epsilon}{4} \left(\varphi^i  \varphi^i\right)^2 \right]
\end{equation}
The IR fixed point of the theory is denoted the Wilson-Fisher point and occurs at $g= \frac{8\pi^2}{N+8}\epsilon+ \cdots$. At the point $g=\epsilon=0$ the theory is free and possesses HS symmetry. As we turn on $g$ the fundamental field and conserved currents acquire anomalous dimensions. These anomalous dimensions have been computed long ago in \cite{Wilson:1973jj} and they are:
\begin{align}
&\gamma_{\varphi^2_S} = \frac{N+2}{N+8}\epsilon+\cdots,~~~~~~~~~~~~~~~~~~~~~~~~~~~\gamma_{\varphi^2_T} = \frac{2}{N+8}\epsilon+\cdots,~~~ \nonumber \cr
&\gamma_\varphi=\frac{N+2}{4(N+8)^2}\epsilon^2+\cdots,
~~~~~~~~~~~~~~~~~~~~~~\gamma_{S,\ell} = 2\gamma_\varphi \left(1-\frac{6}{\ell(\ell+1)} \right)+\cdots,\cr
&\gamma_{T,\ell} = 2\gamma_\varphi \left(1-\frac{N+6}{N+2}\frac{2}{\ell(\ell+1)} \right)+\cdots,~~~\gamma_{A,\ell} = 2\gamma_\varphi \left(1-\frac{2}{\ell(\ell+1)} \right)+\cdots \nonumber
\end{align}
In perfect agreement with our results. 

\subsection{Large $N$ critical $O(N)$ model}

Let us now study a model in $2<d<4$ dimensions with $O(N)$ global symmetry in the limit of large $N$. $1/\sqrt{N}$ will play the role of the small coupling constant. In order to apply our machinery we study the four-point correlator ${\cal G}_{ijkl}(u,v)$ exactly as above. Since $d \neq 4$ it follows that there are no non-trivial solutions to the crossing equations to order $1/\sqrt{N}$ and also that $\gamma_{\varphi}$ vanishes at this order. We can then expand the correlation functions as a series in $1/N$ and study crossing order by order. From the explicit expressions for the correlators at $N=\infty$ we have
\begin{eqnarray} 
{\cal G}_S(u,v) &=& 1 + \frac{1}{N}{\cal G}^{(1)}_S(u,v)  + \cdots\\
{\cal G}_T(u,v) &=& u^{\frac{d-2}{2}}\left(1+\frac{1}{v^{\frac{d-2}{2}}} \right)+ \frac{1}{N}{\cal G}^{(1)}_T(u,v)  + \cdots\\
{\cal G}_A(u,v) &=& u^{\frac{d-2}{2}}\left(1-\frac{1}{v^{\frac{d-2}{2}}} \right) + \frac{1}{N}{\cal G}^{(1)}_A(u,v)  + \cdots
\end{eqnarray}
The intermediate operators at leading order are the identity operator and the symmetric traceless and anti-symmetric conserved currents $J_T^{(\ell)},J_A^{(\ell)}$ with spin even and odd respectively and twist $d-2$. The crossing equations imply
\begin{equation}
{\cal G}^{(1)}_S(u,v)= \left( \frac{u}{v}\right)^{\frac{d-2}{2}}\left( 1 + v^\frac{d-2}{2} +\frac{ {\cal G}^{(1)}_T(v,u)-{\cal G}^{(1)}_A(v,u) }{2}\right) + \gamma_{\varphi}^{(1)}\log \frac{u}{v}
\end{equation}
The operators with the lowest twist exchanged on the r.h.s. of this equations are the currents, of approximate twist $\tau=d-2$. Hence, in the limit $v \to 0$ we obtain
\begin{equation}
{\cal G}^{(1)}_S(u,v)= \left( \frac{u}{v}\right)^{\frac{d-2}{2}}+ \cdots
\end{equation}
where the dots denote regular terms, or which diverge as much as a single conformal block. The intermediate operators contributing to ${\cal G}^{(1)}_S(u,v)$ are the currents transforming in the singlet representation of $O(N)$. From the divergent behaviour above it follows their OPE coefficient is equal to the OPE coefficient of the symmetric traceless representation, times the factor $1/N$. This is valid to all orders in $1/J^2$, where $J^2=(\ell+d/2-2)(\ell+d/2-1)$. In addition, very much as in section 2, the crossing equations are consistent with the exchange of a scalar operator, which we denote by $\sigma$, of classical dimension two and in the singlet representation of $O(N)$. This leads to the term
\begin{equation}
{\cal G}^{(1)}_S(u,v) = a_{\varphi^2\sigma} u^{\frac{\Delta_\sigma}{2}} g_{\Delta_\sigma,0}(u,v) + \cdots
\end{equation}
Note that $\sigma$ enters with its classical dimension, since the OPE coefficient is already of order $1/N$. We cannot fix the parameter $a_{\varphi^2\sigma}$ from crossing alone. $a_{\varphi^2\sigma}=0$ corresponds to the free theory. Let us see the consequences of this term. To order $1/N$ the external operators as well as the currents acquire anomalous dimensions\footnote{Since there are no corrections to order $1/\sqrt{N}$, we denote by $ \gamma^{(1)}_{\varphi} $, etc, the corrections to order $1/N$.}:
\begin{eqnarray}
\Delta_\varphi &=& \frac{d-2}{2} + \frac{1}{N} \gamma^{(1)}_{\varphi} + \cdots\\
\Delta_{J^{(\ell)}} &=& d-2+\ell + \frac{1}{N}  \gamma_{\ell}^{(1)}+ \cdots
\end{eqnarray}
The discussion at order $1/N$ for the symmetric traceless and anti-symmetric currents goes along the same lines as the discussion in section two for $d \neq 4$. The presence of $\sigma$ in the singlet channel produces a correction to the dimensions of the traceless currents, see the crossing equations for $O(N)$ given in (\ref{crossON}). In addition we have the uniform solution, proportional to $\gamma^{(1)}_\varphi$. As a result we obtain
\begin{equation}
\gamma_{T,\ell}^{(1)}=\gamma_{A,\ell}^{(1)} = 2 \gamma^{(1)}_\varphi -\frac{(d-4)^2}{2}\frac{a_{\varphi^2\chi}}{(\ell+\frac{d-2}{2})(\ell+\frac{d-2}{2}-1)} ,
\end{equation}
As for the general discussion in section two, we don't have the freedom to add solutions with finite support in the spin, so that this result is valid for all values of $\ell$. Conservation of the global $O(N)$ symmetry implies $\gamma_1^{(2)}=0$. This gives a relation between $\gamma^{(2)}_\varphi$ and the OPE coefficient $a_{\varphi^2\chi}$ leaving us with a single free parameter. We end up with
\begin{equation}
\label{traceless}
\gamma_{T,\ell}^{(1)}=\gamma_{A,\ell}^{(1)}  = 2 \gamma^{(1)}_\varphi \left(1 - \frac{d(d-2)}{(d+2\ell-2)(d+2\ell-4)} \right) ,
\end{equation}
 The analysis of $1/N$ corrections to the dimension of currents in the singlet representation is more involved, since they appear at the next order in a $1/N$ expansion. In order to analyse them we expand the crossing equations (\ref{crossON}) to order $1/N^2$ and focus in the relevant contributions to ${\cal G}^{(2)}_S(u,v)$. We obtain
\begin{equation}
{\cal G}^{(2)}_S(u,v) = \gamma^{(1)}_\varphi  \log u {\cal G}^{(1)}_S(u,v) + \frac{u^{\frac{d-2}{2}}}{v^{\frac{d-2}{2}}} {\cal G}^{(1)}_S(v,u)  +\frac{u^{\frac{d-2}{2}}}{v^{\frac{d-2}{2}}}  \left({\cal G}^{(2)}_T(v,u)-{\cal G}^{(2)}_A(v,u) \right) + \cdots
\end{equation}
The first two terms lead to exactly the same result as for the traceless representations (\ref{traceless}). The last contribution, which we denote ${\cal G}^{(2)}_{T-A}(v,u)$, is more interesting.  The anomalous dimensions (\ref{traceless}) lead to a contribution proportional to $\log^2v$, which can be exactly computed. This contribution is given by
 \begin{align}
&\left. {\cal G}^{(2)}_{T-A}(v,u)\right|_{\log^2 v} =\\
&~~~~~~~~~~~~~~~ \frac{1}{2} \left( \gamma^{(1)}_\varphi \right)^2\left(H_{T-A,d-2}^{(0)}(v,u) -\frac{d(d-2)}{2} H_{T-A,d-2}^{(1)}(v,u) +\frac{d^2(d-2)^2}{16}H_{T-A,d-2}^{(2)}(v,u)  \right) \nonumber
 \end{align}
 We will furthermore focus in the contribution proportional to $\log u$. As explained in appendix \ref{twistcb} the only twist conformal block that contributes is $H_{T-A,d-2}^{(2)}(v,u)$. The corresponding contribution to the anomalous dimensions of the currents in the singlet representation is then given by
 \begin{equation}
 \label{crossingsinglet}
\left. \sum_\rho B_{\rho} H^{(\rho)}_{d-2}(u,v) \right|_{\log^2 v} = \frac{u^{\frac{d-2}{2}}}{v^{\frac{d-2}{2}}} N \left( \gamma^{(1)}_\varphi \right)^2 \left. \frac{d^2(d-2)^2}{32}H_{T-A,d-2}^{(2)}(v,u) \right|_{\log u}
 \end{equation}
The extra factor of $N$ on the r.h.s. arises because  the OPE coefficients of currents in the singlet representation are suppressed by this factor. The precise structure of divergences can be reproduced provided $\rho = \frac{d-2}{2}+n$, with $n=0,1,\cdots$. From the results in appendix \ref{twistcb} we can construct the relevant twist conformal blocks to arbitrary high orders and solve for $B_{\frac{d-2}{2}+n} =\hat B_{\frac{d-2}{2}+n} N \left( \gamma^{(1)}_\varphi \right)^2$. For the first few cases we obtain
\begin{eqnarray}
\hat B_{\frac{d-2}{2}} = -\frac{2 \pi  \csc \left(\frac{\pi  d}{2}\right) \Gamma \left(\frac{d}{2}+1\right)^2}{d-4} ,~~~\hat B_{\frac{d-2}{2}+1} = \frac{(d-2)(d-4)(d-6)}{24}\hat B_{\frac{d-2}{2}},
\end{eqnarray}
and so on. With enough patience one can compute $B_{\frac{d-2}{2}+n}$ for arbitrarily large $n$, and we have computed many. The computation is simplified by noting that acting with ${\cal D}_{sat}$ on both sides of (\ref{crossingsinglet}) gives zero. This allows to find recursively the coefficients $B_{\frac{d-2}{2}+n}$ once the first two are found. The resulting expression for the anomalous dimensions can be resummed into the following expression
\begin{equation}
\label{criticalsing}
\gamma_{S,\ell}^{(1)}= 2 \gamma^{(1)}_\varphi \left(1 - \frac{d(d-2)}{(d+2\ell-2)(d+2\ell-4)} \right) + 2 N \left( \gamma^{(1)}_\varphi \right)^2 \hat B_{\frac{d-2}{2}} \frac{\Gamma \left(\frac{1}{2} \left(-d+\sqrt{4 J^2+1}+5\right)\right)}{J^2 \Gamma \left(\frac{1}{2} \left(d+\sqrt{4 J^2+1}-3\right)\right)}
\end{equation}
where $J^2=(\ell+d/2-2)(\ell+d/2-1)$. So far we have fixed the anomalous dimensions of the currents of all representations in terms of $\gamma^{(1)}_\varphi$ and $N$. As for the previous section, we don't have to freedom to add solutions with finite support on the spin. Hence, these results are valid for all values of the spin. It turns out we can do even better than this. By requiring conservation of the stress tensor we have $\gamma_{S,2}^{(1)}=0$, which actually imposes an additional constraint. We find
\begin{equation}
 \gamma^{(1)}_\varphi \left(2 \gamma^{(1)}_\varphi  N \hat B_{\frac{d-2}{2}}+ d\, \Gamma(d-1) \right) =0
\end{equation}
One solution is the result for the free theory. The other solution agrees with the known expression for the critical $N$ $O(N)$ model in the large $N$ limit! This model is conveniently described by the following action
\begin{equation}
S = \int d^d x\left[\frac{1}{2} \partial_\mu \varphi^i \partial^\mu \varphi^i+ \frac{1}{2\sqrt{N}} \sigma \varphi^i\varphi^i \right]
\end{equation}
where $\sigma$ is an auxiliary scalar field of classical dimension $\Delta_\sigma = 2+\cdots$. The anomalous dimensions for all the currents have also been computed \cite{Lang:1992zw} and they perfectly agree with our results.  

Before concluding with this example, we would like to make the following remark. The method of large spin perturbation theory gives the anomalous dimensions as a series expansion in $1/J^2$, sometimes asymptotic. In order to re-sum this series unambiguously, as in (\ref{criticalsing}) we need to provide extra information. The claim is that  (\ref{criticalsing}) is the unique function with the correct asymptotics and which is analytic in the half-plane $4J^2-1>(d-3)^2$. Note that this excludes the point $\ell=0$. This is actually quite nice, because in the critical model the operator $\varphi^i \varphi^i$, scalar, singlet of $O(N)$ and of classical dimension $d-2$, is absent. In the free theory $\gamma^{(1)}_\varphi =0$ and this constraint disappears, as it should.

\subsection{Cubic models in $d=6-\epsilon$}
Let us now consider models in six dimensions. Let us start with a single fundamental field $\varphi$. In this case $\varphi$ has dimension $2$ at the HS symmetric point, and crossing symmetry admits solutions where $\varphi$ is present in the OPE $\varphi \times \varphi$. Our discussion in section \ref{genericmodel} applies and the anomalous dimensions of the broken currents is exactly given by (\ref{sigmaexchange}), with $d=6$ and the OPE coefficient replaced by $a_{\varphi\varphi\varphi}$. Conservation of the stress tensor implies $\gamma_2=0$ and we arrive to the final result
\begin{equation}
\label{cubic}
\gamma_\ell^{(2)} = 2 \gamma_{\varphi}^{(2)} \left(1 -\frac{12}{(\ell+2)(\ell+1)} \right)
\end{equation}
This exactly agrees with the known result for the theory of a single scalar and cubic Lagrangian in $d=6-\epsilon$
\begin{equation}
S = \int d^d x \left[\frac{1}{2}(\partial \varphi)^2+ \frac{g}{3!} \varphi^3 \right]
\end{equation}
This model has a non-unitary IR fixed point and its continuation to $\epsilon \to 4$ describes the Lee-Yang edge singularity in the Ising model \cite{Fisher:1978pf}. Note that provided we are close to a HS symmetric point we can apply our method, without assuming the model is unitary. 

We could also consider a model of $N$ scalar fields $\varphi^i$ with $O(N)$ symmetry, coupled to a singlet scalar field $\sigma$. In six dimensions also a cubic potential for $\sigma$ is allowed. An example of this is the cubic model in $d=6-\epsilon$ dimensions and $O(N)$-invariant cubic interactions \cite{Fei:2014yja}
\begin{equation}
S = \int d^d x \left[\frac{1}{2}(\partial \varphi)^2+ \frac{1}{2}(\partial \sigma)^2 + \frac{g_1}{2}\sigma \varphi^i \varphi^i + \frac{g_2}{6}\sigma^3 \right]
\end{equation}
For large enough $N$ pertubative fixed points exist with $g_1,g_2 \sim \epsilon^{1/2}$. The free theory contains two independent towers of $O(N)-$singlet conserved currents, of the schematic form
\begin{eqnarray}
J^{(\ell)}_\phi = \varphi^i \partial_{\mu_1} \cdots\partial_{\mu_s} \varphi^i\\
J^{(\ell)}_\sigma = \sigma \partial_{\mu_1} \cdots\partial_{\mu_s} \sigma
\end{eqnarray}
The interaction term proportional to $g_1$ induces a mixing between these currents. In addition, we have the usual symmetric traceless and anti-symmetric currents. In order to apply our formalism we can again consider the four point correlator of fields $\varphi$. Most of our previous discussion goes through and can be performed directly in six dimensions. At order $g \sim g_1,g_2$ we find that the only allowed solution to crossing takes the form
\begin{equation}
\gamma^{(1)}_{T,\ell} =\gamma^{(1)}_{A,\ell} = 2 \gamma^{(1)}_{\varphi},
\end{equation}
Conservation of the global $O(N)$ symmetry then implies $\gamma^{(1)}_{A,1}=0$ and hence the above anomalous dimensions vanish at order $g$. At order $g^2$ crossing symmetry allows the presence of a new scalar operator of classical dimension two in the singlet channel. This leads to a contribution with exactly the form (\ref{sigmaexchange}) in $d=6$:
\begin{equation}
\gamma_{T,\ell}^{(2)}= \gamma_{A,\ell}^{(2)} = 2 \gamma^{(2)}_\varphi -2\frac{a_{\varphi^2\sigma}}{(\ell+2)(\ell+1)} 
\end{equation}
Conservation of the global current then leads to a further constraint and to the final result
\begin{equation}
\label{cubicon}
\gamma_{T,\ell}^{(2)}= \gamma_{A,\ell}^{(2)} = 2 \gamma^{(2)}_\varphi\left(1 -\frac{6}{(\ell+2)(\ell+1)} \right)
\end{equation}
Which is in perfect agreement with the result for the cubic model, see \cite{Giombi:2016hkj}. Degeneracy makes the discussion of the singlet currents a bit more complicated. One should apply the procedure used here to mixed correlators involving also the field $\sigma$. We will not do this here, but we don't expect any obstacles. 

Before concluding this section, note an interesting point. The discussions that led to (\ref{cubic}) and (\ref{cubicon}) are almost identical, except in the first case we require conservation of the stress tensor, while in the second conservation of the global currents, hence the different factors in the numerator. 

\section{A simple model of weakly coupled gauge theory}
\label{gaugetheories}
In this section we will discuss a scalar model in $d$-dimensions with weakly broken higher spin symmetry, but such that the fundamental field $\varphi$ is not part of the spectrum. An example of this would be a sector of weakly coupled gauge theories. In order to apply the formalism developed in \cite{Alday:2016njk} we consider the four point correlator of $J^{(0)}$. For definiteness we can think of an almost free theory, where $J^{(0)} = \varphi^2$. Let us assume for simplicity that $\varphi^2$  does not receive corrections. The extension to the case in which $\varphi^2$ is not protected is straightforward. The four point function is
\begin{equation}
\langle \varphi^2(x_1)\varphi^2(x_2)\varphi^2(x_3)\varphi^2(x_4) \rangle = \frac{{\cal G}(u,v)}{x_{12}^{2(d-2)}x_{34}^{2(d-2)}}
\end{equation}     
Crossing symmetry implies $v^{d-2}{\cal G}(u,v)=u^{d-2}{\cal G}(v,u)$. The explicit result in the free theory can be found in \cite{Dolan:2000ut}. At the HS symmetric point we can decompose this correlator in twist conformal blocks 
\begin{equation}
{\cal G}(u,v)=1 + H^{(0)}_{d-2}(u,v)+...,~~~~~~H^{(0)}_{d-2}(u,v) = \frac{1}{c} \left(\left( \frac{u}{v}\right)^{\frac{d-2}{2}} + u^{\frac{d-2}{2}} \right)
\end{equation}   
where $c$ is proportional to the central charge of the theory and we have kept only the leading twist contribution, which arises from the HS conserved currents.  Note that the form of $H^{(0)}_{d-2}(u,v)$ is exactly as before, up to an overall factor. Consequently, the OPE coefficients $a^{(0)}_\ell$ also agree with (\ref{HSope}) times $1/c$. Note that crossing symmetry maps the divergent part of $H^{(0)}_{d-2}(u,v)$ to itself. As discussed in \cite{Alday:2015ota,Alday:2013cwa}, this is still true as we turn on the coupling. We will revisit the discussion of \cite{Alday:2015ota,Alday:2013cwa} in view of the technology developed here. As we turn on the coupling the conserved currents acquire anomalous dimensions.  In the direct channel we can study the spectrum by focusing in the piece proportional to $\log u$. On the other hand, in the dual channel we can study the spectrum by focusing in the piece proportional to $\log v$. This leads to the following relation
\begin{equation}
v^{d-2}\sum B_\rho \left. H^{(\rho)}_{d-2}(u,v)\right|_{\log v}   = u^{d-2}\sum B_\rho \left. H^{(\rho)}_{d-2}(v,u)\right|_{\log u} 
\end{equation}
which arises from the crossing equation projected over the piece proportional to $\log u \log v$ in a small $u,v$ expansion, and the equation is understood to include only the divergent contributions. The coefficients $B_\rho$ encode the large spin expansion of the anomalous dimensions. First we note the following simple fact: the small $v$ behaviour of the r.h.s implies a divergence $v^{-\frac{d-2}{2}} \log v$. As discussed in appendix A.2 the presence of this divergence implies the presence of the twist conformal block $H^{(0,\log J)}_{d-2}(u,v)$, with a $\log J$ insertion, on the l.h.s. and hence a logarithmic behaviour for the anomalous dimensions. Furthermore, the twist conformal blocks $H^{(m,\log J)}_{d-2}(u,v)$ behave as 
\begin{equation}
H^{(m,\log J)}_{d-2}(u,v) \sim \frac{u^{\frac{d-2}{2}}}{v^{\frac{d-2}{2}-m}}(\sum_{i,j=0}\alpha_{ij}u^i v^j+ \sum_{i,j=0}\beta_{ij}u^i v^j \log v)
\end{equation}
For $m=0$ only the first term for small $v$ contains a $\log v$. We see that the term with $m=0$ is allowed and will give a crossing symmetric contribution. On the other hand, none of the terms with $m>0$ will. So that the anomalous dimensions contain a logarithmic divergent piece, but no term going like $1/J^n\log J$, at first order in the coupling. 

Let us now focus in the problem in $d=4$. We will use holomorphic variables, where the divergent contribution to twist conformal blocks factorises. The divergent  contribution at zero order from the conserved currents is given by
\begin{equation}
H^{(0)} _\tau(z,\bar z)= \frac{z^{\tau/2}}{\bar z-z} F_{\frac{\tau-2}{2}}(z) \bar H_\tau^{(0)}(\bar z)
\end{equation}  
where $\tau=2$. Following \cite{Alday:2015eya,Alday:2015ewa} it is convenient to define new OPE coefficients given by
\begin{equation}
a_{\ell} =a^{(0)}_{\ell+\gamma_\ell/2}(1+\frac{1}{2} \partial_\ell \gamma_\ell) \hat a_{\ell}
\end{equation}
As we turn on the coupling there are two kind of contributions. One arising from corrections to the OPE coefficients $\hat a_{\ell}$, and the other arising from the anomalous dimensions $\gamma_\ell$. As shown in the appendix A to \cite{Alday:2015ewa}, the corrections to the dimensions have the effect of shifting the factor $\tau$ in $z^{\tau/2}F_{\frac{\tau-2}{2}}$ in the expression for $H^{(0)} _2(z,\bar z)$, while keeping the $\tau$ index in $\bar{H}_\tau^{(0)}(\bar z)$ untouched.\footnote{In  \cite{Alday:2015ewa} this was proven to first order in $z$. The proof to all orders in $z$ goes in exactly the same way. The basic reason is that the divergent factor $\bar{H}_\tau^{(0)}(\bar z)$ can be decomposed into anti-holomorphic  conformal blocks depending only on $J$ and not $\ell$ and $\tau$ separately, so that the dependence on the anomalous dimension can be reabsorved into the proper definition of $J$. See appendix A  to \cite{Alday:2015ewa}.}
Setting $\tau=d-2+\gamma$ we obtain:
\begin{equation}
z^{\tau/2} F_{\frac{\tau-2}{2}}(z) = z\left(1 + \gamma \left(\frac{1}{2} \log z - \frac{1}{4} \log(1-z)\right) + \cdots \right)
\end{equation}
We then assume that the OPE coefficients $ \hat a_{\ell}$ and the anomalous dimensions $\gamma_\ell$ admit an expansion (were also $\log J$ terms are allowed)
\begin{equation}
\hat a_{\ell} = 1+g \sum_\rho \frac{A_\rho}{J^{2\rho}},~~~~~\gamma_\ell = \sum_\rho \frac{B_\rho}{J^{2\rho}}
\end{equation}
This leads to the following contribution to ${\cal G}^{(1)}(z,\bar z)$ from the HS (now broken) currents  
\begin{equation}
\left. {\cal G}^{(1)}(z,\bar z) \right|_{\tau=2}= \frac{z}{\bar z-z} \left( \sum_\rho A_\rho \bar H_{d-2}^{(\rho)}(\bar z) +  \left(\frac{1}{2} \log z - \frac{1}{4} \log(1-z)\right) \sum_\rho B_\rho \bar H_{d-2}^{(\rho)}(\bar z)\right)
\end{equation}%
where only terms with enhanced divergence respect to a single conformal block are kept. We can write the crossing symmetry equation as
\begin{eqnarray}
 \frac{(1-z)^2}{z} \left( \sum_\rho A_\rho \bar H_{d-2}^{(\rho)}(\bar z) +  \left(\frac{1}{2} \log z - \frac{1}{4} \log(1-z)\right) \sum_\rho B_\rho \bar H_{d-2}^{(\rho)}(\bar z)\right) = (z \leftrightarrow 1-\bar z)
\end{eqnarray}%
This should be interpreted as an equation for the coefficients $A_\rho,B_\rho$. Now we make the following observation. The r.h.s contains a divergence $\frac{\log (1-\bar z)}{(1-\bar z)}$. This implies the presence of $\rho=(0,\log J)$ on the l.h.s., see appendix \ref{twistcb} . On the other hand, this generates a $\log^2(1-\bar z)$ term which is absent on the r.h.s. This term should be canceled by the appropriate linear combination of $H_\tau^{(\rho)}(\bar z)$ with $\rho=1,2,3,\cdots$. In other words, we need to consider
\begin{equation}
\bar H_{d-2}^{(0,\log J)}(\bar z) + \sum_{m=0} B_m \bar H_{d-2}^{(m)}(\bar z)
\end{equation}
and this should not contain a $\log^2(1-\bar z)$ piece. From the results in appendix \ref{twistcb} we can solve iteratively. We find
\begin{equation}
B_0 = \gamma_e,~~~B_1= \frac{1}{6},~~~B_2 = -\frac{1}{30},~~~B_3=\frac{4}{315},~~~B_4 = -\frac{1}{105}, 
\end{equation}
and so on. The attentive reader will notice that these are the coefficients of the expansion of $S_1(\ell)$ in inverse powers of $J^2=\ell(\ell+1)$ where $S_1$ is the harmonic number! In this case
\begin{equation}
\label{harmonic}
\bar H_{d-2}^{(0,\log J)}(\bar z) + \sum_{m=0} B_m \bar H_{d-2}^{(m)}(\bar z) = -\frac{1}{2} \frac{\log(1-\bar z)}{(1-\bar z)}
\end{equation}
which indeed is consistent with crossing symmetry. In addition, we can have a uniform contribution. A similar reasoning applies for corrections to the OPE coefficients $\hat a^{(1)}_\ell$. In summary, crossing symmetry for the present models leads to a correction of the form
\begin{eqnarray}
\gamma_\ell^{(1)} = c_1 + c_2 S_1(\ell)\\
\hat a^{(1)}_\ell = c_3 -c_1 S_1(\ell)
\end{eqnarray}
where $c_{1,2,3}$ are numeric constants. This follows just from the analysis of the enhanced divergent terms. We have studied finite support solutions in this case. Although there are truncated solutions in the spin, none of them contributes to the leading twist operators. As a result, the results above should be valid also for finite values of the spin. Written in terms of the usual OPE coefficients $a_\ell =a^{(0)}_\ell\left(1+ g a^{(1)}_\ell + \cdots \right) $ the corrections above take the form
\begin{eqnarray}
\label{final}
\gamma_\ell^{(1)} &=& c_1 + c_2 S_1(\ell)\\
a^{(1)}_\ell &=& c_3+\frac{\zeta_2}{2}c_2+ c_2 S_1(\ell)^2 - c_1 S_1(2\ell) -c_2 S_1(\ell)S_1(2\ell) - \frac{c_2}{2} S_2(\ell)
\end{eqnarray}
So that we have fixed the whole spectrum of broken currents, and their OPE coefficients, in terms of three numerical constants. There is still two more constraints we should impose, namely
\begin{equation}
\gamma_2^{(1)}=0,~~~~~~a_0=   \frac{d^2 \Delta_{\varphi^2}^2}{(d-1)^2 c_T}
\end{equation}
The first relation is conservation of the stress tensor. The second arises from the known expression for the three point function between two $J_0$ and the stress tensor: it is known that this OPE is universal and depends only on the dimension of the operator and the central charge $c_T$. Note that the central charge of the theory will enter through the second constraint. For the specific case at hand $d=4$.  This allows to fully write the answer, for all values of the spin, in terms of a single parameter! which has the interpretation of the coupling constant. 

Let us compare our results to the known results for ${\cal N}=4$ SYM. This theory has an $SO(6)$ $R-$symmetry group and the scalars $\varphi^I$ transform in the fundamental representation of $SO(6)$. The bilinear currents made out of the fundamental field decompose again into the singlet, symmetric traceless and anti-symmetric representations. ${\cal N}=4$ SYM also contains fermions and gauge bosons. It turns out the currents transforming in the singlet and the anti-symmetric representation mix with currents made out of fermions and gauge bosons. On the other hand, currents in the symmetric traceless representation
\begin{equation}
J_{{\bf 20'}}^{(\ell)} = Tr \varphi^{(I} \partial_{\mu_1} \cdots  \partial_{\mu_\ell}  \varphi^{J)} 
\end{equation}
are non degenerate. To the order we have worked our results should apply to these currents. The current is BPS for $\ell=0$, since then it is the super-conformal  primary of the stress-tensor multiplet. This implies $c_1=0$ above. In addition, the three point function of three BPS operators is also protected, so that $a^{(1)}_0=0$. This leads to the final answer
\begin{eqnarray}
\gamma_\ell^{(1)} &=& c_2 S_1(\ell)\\
a^{(1)}_\ell &=& \frac{c_2}{2} \left( 2 S_1(\ell) (S_1(\ell)- S_1(2\ell))-S_2(\ell)\right)
\end{eqnarray}
Which precisely agrees with the known results! (see \cite{Plefka:2012rd} for the computation of the OPE coefficient). 

\section{Conclusions}
In the present paper we have studied conformal field theories with weakly broken HS symmetry. At the HS symmetric point we assume the existence of a fundamental field satisfying the free equations of motion $\partial_\mu \partial^\mu \varphi=0$ and a tower of conserved currents $J^{(\ell)}$. As we break HS symmetry these currents and the fundamental field acquire an anomalous dimension. We consider the four-point correlator of the fundamental field and find the spectrum, to all values of the spin, consistent with crossing symmetry and the presence of a conserved stress tensor. Our method uses only the symmetries of the problem. After studying a generic model with a single scalar field, we consider several examples with global symmetry, in various dimensions, to rediscover the spectrum of several Lagrangian systems, although no Lagrangian was assumed. We then study models where the fundamental field is not part of the spectrum. This case is relevant to study weakly coupled gauge theories. The anomalous dimensions of the broken currents have a different behaviour, involving Harmonic numbers. In all cases crossing symmetry fixes the spectrum of broken currents to all values of the spin, in terms of a single parameter, which can be though of as the coupling constant. Our method provides an alternative to usual perturbation theory and does not rely on a Lagrangian description. 

There are several directions one can pursue. The most obvious one is to generalise the results of this paper to higher orders in the breaking parameter $g$.  As we do this, new operators will enter in the OPE $\varphi \times \varphi$ and the treatment becomes in principle more complicated, although we expect this to be a technical obstacle, not conceptual. At the order we have considered in this paper the appearance of such operators is not relevant. It would be interesting to develop a set of diagrammatic rules that allow to compute the spectrum to higher orders. 

It would be interesting to consider two-dimensional models with HS symmetry. An interesting example would be the non-linear sigma model. In this case $d-2 \to 0$ and we expect the power law behaviour with the spin to become logarithmic. Indeed this agrees with the explicit results \cite{Giombi:2016hkj}. One may be able to understand this case by carefully considering a limit of the cases studied here, but the limit may be subtle. Another interesting extension would be to consider almost free fermions. This would allow to study, for instance, the Gross-Neveu and related models. 

It would also be interesting to understand what more can be said about weakly coupled gauge theories. In this case there is no fundamental field but there is still a tower of almost conserved currents. Conformal gauge theories have been studied in the past by similar methods \cite{Alday:2013cwa,Alday:2015eya,Alday:2015ota}, and we have shown in this paper that the methods of \cite{Alday:2016njk} generalise these results to all values of the spin. We find it remarkable that both the spectrum and OPE coefficients of leading twist operators can be completely fixed from crossing symmetry (with the free parameter having the interpretation of the coupling). The method proposed here is a gauge invariant on-shell way to study gauge theories and this can be explored much further. In this paper we have focused in a four-dimensional example with a single scalar field, but there are very interesting examples in other dimensions. Interesting examples are susy gauge theories, such as ${\cal N}=4$ SYM, ABJM, etc, and Chern-Simons gauge theories coupled to matter \cite{Giombi:2011kc,Aharony:2011jz}, with or without supersymmetry. Some explicit results are known \cite{Giombi:2016zwa}.  

It would be interesting to connect our methods to alternative bootstrap-inspired ways to obtain analytic information about the spectrum of interacting CFT. An example is the method developed in \cite{Gopakumar:2016wkt,Gopakumar:2016cpb}, based in the Mellin representation of correlators. It would be very interesting to combine the methods used in the present paper with the methods in those papers, for theories where the assumptions of  \cite{Gopakumar:2016wkt,Gopakumar:2016cpb} hold. It is also tantalising to try to study sectors with large global quantum numbers with our methods. Certain universal semi-classical picture has been argued to arise in this limit, see {\it e.g.}   \cite{Hellerman:2015nra,Monin:2016jmo,Alvarez-Gaume:2016vff}. These papers rely in the Lagrangian formulation of the theory, while our methods may provide a non-Lagrangian way of understanding the same physics.  Another CFT method was developed in \cite{Rychkov:2015naa}. This method relies on the equations of motion, but is very powerful to first order in $\epsilon$. On the other hand, the method presented here allows to "guess" the equations of motions and in principle extends to all orders in $\epsilon$, but has not been used to study operators of the form $\varphi^n$, with $n>2$. It would be interesting to combine both methods. 

It would be interesting to use the method proposed in this paper to look for new weakly coupled CFT's or give a complete classification of those. Our results suggest (and prove in many cases) that there is a one-to-one correspondence between solutions to the crossing equations around HS symmetric points and weakly coupled Lagrangians. This is a generalisation of the results of \cite{Heemskerk:2009pn} to weakly coupled theories. It would be interesting to consider systems with more fields and more general global symmetries. 

\section*{Acknowledgements} 
We are grateful to S. Giombi, Z. Komargodski, T. Lukowski and A. Zhiboedov for useful discussions and comments on the manuscript. 
This work was supported by ERC STG grant 306260. The author is a Wolfson Royal Society Research Merit Award holder.

\appendix

\section{Twist conformal blocks}
\label{twistcb}
In this appendix we give useful results regarding the twist conformal blocks $H_{\tau}^{(m)}(u,v)$ relevant for this paper. By definition, the twist conformal blocks capture the contribution from a given twist to a four point correlator with an additional spin-dependent insertion for $m \neq 0$. They admit the following decomposition in conformal blocks
\begin{equation}
H_{\tau}^{(m)}(u,v)=\sum_{\ell} a^{(0)}_{\tau,\ell} J^{-2m} u^{\frac{\tau}{2}} g_{\tau,\ell}(u,v)
\end{equation}
where $J^2=(\ell+\tau/2)(\ell+\tau/2-1)$ and $a^{(0)}_{\tau,\ell}$ are the OPE coefficients at the point of large twist degeneracy. In the present paper this is the HS symmetric point. We will be concerned with the twist at the saturation point $\tau=d-2$. In this case 
\begin{equation}
{\cal D}_{sat} H_{d-2}^{(m)}(u,v) = 0
\end{equation}
where ${\cal D}_{sat}$ is defined in (\ref{Dsat}). In addition, the Casimir operator ${\cal C}$, connects the sequence of functions
\begin{equation}
\label{recc}
{\cal C} H_{d-2}^{(m+1)}(u,v) = H_{d-2}^{(m)}(u, v) 
\end{equation}
where ${\cal C}$ is given by
\begin{equation} 
{\cal C} = D+ \bar D+(d-2) \frac{z \bar z}{z-\bar z}\left((1-z)\partial - (1-\bar z) \bar \partial \right) +\frac{\tau(2d-\tau-2)}{4}
\end{equation}
where $D=(1-z)z^2\partial^2-z^2 \partial,~\bar D=(1-\bar z)\bar z^2 \bar \partial^2-\bar z^2 \bar \partial$. For most of this paper
\begin{equation}
H_{d-2}^{(0)}(u, v) = u^{\frac{d-2}{2}}+ \left(\frac{u}{v} \right)^{\frac{d-2}{2}}
\end{equation}
This is fixed by HS symmetry. From these three relations, we will construct the functions $H_{d-2}^{(m)}(u,v) $ in several cases. In some of the expressions below it will be convenient to use the set of cross-ratios $(z,\bar z)$, related to the set $(u,v)$ by $z \bar z=u,(1-z)(1-\bar z)=v$. The small $u,v$ limit is chosen to be mapped to the small $z,1-\bar z$ limit. We hope the use of two sets of cross-ratios will not cause confusion. 

\subsection{$H_{d-2}^{(m)}(u,v)$ in $d-$dimensions}
Starting with the twist conformal block $H_{d-2}^{(0)}(u,v)$ we use the recurrence relation to construct the sequence of functions $H_{d-2}^{(m)}(u,v)$. We will analytically continue in the parameter $d$. For several applications we will be interested in the singular behaviour of $H_{d-2}^{(m)}(u,v)$, captured by the expansion
\begin{equation}
\label{smallvexp}
H_{d-2}^{(m)}(u,v)=\frac{1}{v^{\frac{d-2}{2}-m}} \left(h^{(m)}_0(u)+h^{(m)}_1(u) v +h^{(m)}_2(u) v^2 + \cdots \right)
\end{equation}
Plugging this expansion into the recurrence relation (\ref{recc}) we can actually solve for all the functions $h^{(m)}_i(u)$. For instance
\begin{equation}
h^{(m+1)}_0(u) =4 \frac{h^{(m)}_0(u) }{(1-u)(2m+4-d)^2}
\end{equation}
Together with the known value for $h^{(0)}_0(u)$ this leads to
\begin{equation}
 h^{(m)}_0(u) = \frac{u^{\frac{d-2}{2}}}{(1-u)^m}\frac{\Gamma^2\left(\frac{d-2}{2} -m\right)}{\Gamma^2\left(\frac{d-2}{2} \right)}
\end{equation}
and so on. An important feature is that already $H_{d-2}^{(1)}(u,v)$ contains the whole tower of divergent terms in (\ref{smallvexp}). The above solution is generic. In the particular case $\frac{d-2}{2} -m \in Z$ we may develop divergent terms which behave as $\log^2 v$. This may happen if $m$ is integer and for even $d$. Below we will explore in detail the case $d=4$. Another interesting case involves non-integer values of $d,m$. For us the following case will be relevant

\begin{equation}
H_{d-2}^{(\frac{d-2}{2}+n)}(z,\bar z)=q^{(n)}(z,\bar z) \log^2(1-\bar z)
\end{equation}
where $n=0,1,\cdots$ and we are disregarding terms without an enhanced divergence respect to a single conformal block. The functions $q^{(n)}(z,\bar z)$ have a regular expansion around $z=0,\bar z=1$ and satisfy
\begin{equation}
{\cal D}_{sat} q^{(n)}(z,\bar z) =0,~~~~~{\cal C}q^{(n+1)}(z,\bar z)=q^{(n)}(z,\bar z)
\end{equation}
This should be supplemented with the correct boundary conditions. Let us start with the limit $\bar z \to 1$. We can determine
\begin{equation}
{\cal C}^m\left(\log^2(1-\bar z) h(z) \right) = 2 \frac{\Gamma^2(m)h(z)}{(1-\bar z)^m} + \cdots
\end{equation}
This leads to the following boundary condition for $q^{(0)}(z,\bar z)$
\begin{equation}
q^{(0)}(z,\bar z) = \frac{1}{2\Gamma^2\left(\frac{d-2}{2}\right)} \left(\frac{z}{1-z}\right)^{\frac{d-2}{2}} + \cdots
\end{equation}
as $\bar z \to 1$. In general $q^{(n)}(z,\bar z) \sim (1-\bar z)^n$. Furthermore, we need to find $q^{(0)}(z,\bar z)$ as $z \to 0$. This can be done in different ways. One way is by direct computation using the explicit form of collinear conformal blocks. Another way is taking carefully the limit $m \to \frac{d-2}{2}$ in the expression (\ref{smallvexp}). As we take the limit the power law divergence disappears, leaving behind a $\log^2(1-\bar z)$ divergence. The final answer admits the following expansion around $\bar z=1$
\begin{equation}
q^{(0)}(z,\bar z) = \frac{z^{\frac{d-2}{2}}}{2\Gamma^2\left(\frac{d-2}{2}\right)} \left(1+ \frac{(2-d)(4-d)(6-d)}{24}(1-\bar z)+\cdots \right) + \cdots
\end{equation}
where only the first two terms are shown (but we can compute any arbitrary number of them). With this information we can completely build the functions $q^{(m)}(z,\bar z)$ to any desired order. 
\subsection{Twist conformal blocks with logarithmic insertions}
For application to gauge theories, or more generally theories with HS symmetry without a fundamental field, it is useful to consider twist conformal blocks with logarithmic insertions of the type:
\begin{equation}
H_{d-2}^{(m, \log J)}(u,v)=\sum_{\ell} a^{(0)}_\ell \frac{\log J}{J^m} u^{\frac{d-2}{2}} g_{d-2,\ell}(u,v)
\end{equation}
This object can be easily computed with the following trick. Above we have explained how to compute $H_{d-2}^{(m)}(u,v)$ analiticaly continued in the parameter $m$ as a series around $v=0$. Having this we can compute $H_{d-2}^{(m, \log J)}(u,v)$ by taking appropriate derivatives with respect to $m$, and then setting $m$ to the required valued. Note that by taking more derivatives, we can even compute higher insertions of $\log J$, which will be relevant at higher loops. For example, for the simplest case with a logarithmic insertion we obtain
\begin{equation}
H_{d-2}^{(0, \log J)}(u,v)= -\frac{1}{2}\left(\frac{u}{v}\right)^{\frac{d-2}{2}} \left( \log \frac{v}{1-u} -2 \psi ^{(0)}\left(2-\frac{d}{2}\right)+\frac{2 ((10-3 d) u+2)}{3 (d-4)^2 (u-1)^2} v + \cdots \right)
\end{equation}
We observe a very interesting property: this twist conformal block contains divergences of the type $\frac{\log v}{v^p}$. Whenever crossing symmetry requires the presence of such divergences, this signals the need for a logarithmic behaviour in the anomalous dimensions of intermediate operators. In general  $H_{d-2}^{(m, \log J)}(u,v) \sim \frac{\log v}{v^{\frac{d-2}{2}-m}}$. A important property of $H_{d-2}^{(0, \log J)}(u,v)$  is that $\log v$ appears only in the leading term in a small $v$ expansion. This is not so for $H_{d-2}^{(m, \log J)}(u,v)$ with $m>1$. 
 
\subsection{4d twist conformal blocks}
In this appendix we construct the twist conformal blocks $H_\tau^{(0)}(u,v)$ together with the sequence of functions $H_\tau^{(m)}(u,v)$ directly in four dimensions. Let us recall
\begin{equation}
H_2^{(0)}(u,v)=\frac{u}{v}+u
\end{equation}
In order to study the sequence of functions $H_\tau^{(m)}(u,v)$, it is convenient to use the set of holomorphic cross ratios $(z,\bar z)$. We will be interested in the divergent contribution to $H_\tau^{(m)}(z,\bar z)$ as $\bar z \to 1$, to all orders in $1-\bar z$.  We will keep $\tau$ general, as four dimensional twist conformal blocks display interesting factorisation properties we want to show. The explicit expression for four-dimensional conformal blocks, plus the fact that divergent terms can arise only when summing over the spin, leads to the following factorised form
\begin{equation}
H_{\tau}^{(0)}(z,\bar z)= \frac{1}{\bar z-z} z^{\tau/2}F_{\frac{\tau-2}{2}}(z) \bar H^{(0)}_{\tau}(\bar z)
\end{equation}
very much like in two dimensions! this captures all divergent terms. Twist conformal blocks with $\tau>d-2$ arise if we consider correlators of operators of the form $\varphi^p$. For $p >1$, we need to consider $H_{\tau}^{(0)}(u,v)$ with $\tau=4,6,\cdots$. We have the following relation

\begin{equation}
\sum_{\tau=4}^\infty H_{\tau}^{(0)}(u,v) = \sum_{L=4,6,\cdots}^{2p} a_L \left(\frac{u}{v} \right)^{\frac{L}{2}}~_2F_1(-L/2,-L/2,1;v)
\end{equation}
The r.h.s. arises from the explicit result of the correlator in the HS symmetric point. The functions $\bar H^{(0)}_{\tau}(\bar z)$ can be systematically worked out. The general structure is as follows
\begin{equation}
\bar H^{(0)}_{\tau}(\bar z) = \frac{a_\tau}{(1-\bar z)^{\min(\tau/2,p)}}\left( 1 + b_\tau (1-\bar z) + \cdots\right)
\end{equation}
where the dots include only divergent contributions. For the simplest non-trivial case $p=2$ we obtain
\begin{equation}
\bar H^{(0)}_{\tau}(\bar z) = \frac{a_\tau}{(1-\bar z)^{2}} \left(1+ b_\tau(1-\bar z) \right)
\end{equation}
where
\begin{equation}
a_\tau=\frac{\sqrt{\pi } 2^{4-\tau } \Gamma \left(\frac{\tau }{2}-1\right)}{\Gamma \left(\frac{\tau }{2}-\frac{3}{2}\right)},~~~b_\tau =-\frac{\tau ^2}{4}+\frac{3 \tau }{2}+4 (-1)^{\tau /2}-5
\end{equation}
\subsubsection{Sequence of functions}
Let us now turn to the divergent behaviour of the sequence of functions $H_{\tau}^{(m)}(z,\bar z)$ for $m>0$. As usual, they can be computed by the recurrence relation (\ref{recc}). For the same reasons above, the sequence of functions has the same factorisation properties
\begin{equation}
H_{\tau}^{(m)}(z,\bar z)= \frac{1}{\bar z- z} z^{\tau/2}F_{\frac{\tau-2}{2}}(z) \bar H^{(m)}_{\tau}(\bar z)
\end{equation}
The recurrence relation then leads to 
\begin{equation}
\bar D_{4d} \bar H^{(m+1)}_{\tau}(\bar z) = \bar H^{(m)}_{\tau}(\bar z),~~~\bar D_{4d}=  \bar z \bar D \bar z^{-1}
\end{equation}
which makes the problem again effectively two dimensional. Note that the operator acting on the l.h.s. is slightly different to the operator in the two dimensional case, but again the twist cancels. The dependence on the twist will enter only through the boundary conditions. For $\tau=2$ and the first few values of $m$ we have
\begin{equation}
\bar H^{(0)}_{2}(\bar z) = \frac{1}{1-\bar z},~~~\bar H^{(1)}_{2}(\bar z) = \frac{\bar z}{2} \log^2(1-\bar z),~~~\bar H^{(2)}_{2}(\bar z) =-\frac{1}{2} \bar z \log \bar z \log^2(1-\bar z),
\end{equation}
where we are disregarding non-enhanced terms. The general structure for $m>0$ is as follows
\begin{equation}
\bar{H}^{(m)}_{2}(\bar z) = h^{(m)}_2(\bar z) \log^2(1-\bar z),~~~\bar D_{4d} h^{(m+1)}_2(\bar z) = h^{(m)}_2(\bar z) 
\end{equation}
where $ h^{(m)}_2(\bar z)$ admits an analytic expansion around $\bar z=1$ with $ h^{(m)}_2(\bar z) \sim (1-\bar z)^{m-1}$. The structure for higher values of the twist $\tau$ is very similar. For the case $p=2$ the boundary conditions depend mildly on the twist. In this case we have 
\begin{eqnarray}
\bar H^{(0)}_{\tau}(\bar z) &=& \frac{a_\tau}{(1-\bar z)^{2}} \left(1+ b_\tau(1-\bar z) \right)\\
\bar H^{(1)}_{\tau}(\bar z) &=& \frac{a_\tau}{(1-\bar z)}+ \frac{1}{2}(3a_\tau+b_\tau)\bar z \log^2(1-\bar z)\\
\bar H^{(m)}_{\tau}(\bar z) &=&  (3a_\tau+b_\tau)h^{(m)}(\bar z) \log^2(1-\bar z)~~~ \text{for $m>1$}
\end{eqnarray}
with $\bar D_{4d} h^{(m+1)}(\bar z) = h^{(m)}(\bar z)$ and $ h^{(m)}_\tau(\bar z)$. The dependence on the spin factors out completely! very much as for the two dimensional case.  

\subsection{Logarithmic insertions in 4d}
We can also study logarithmic insertions in twist conformal blocks directly in four dimensions. For $m \geq 1$ we have the following form
\begin{equation}
\bar H^{(m)}_{2}(\bar z) =\bar q^{(m)}(\bar z)  \log^2(1-\bar z),~~~~~\bar q^{(1)}(\bar z) =  \frac{\bar z}{2}
\end{equation}
With $q^{(m)}(\bar z) \sim (1-\bar z)^{m-1}$ as $\bar z \to 1$. We can then use the recursion relations to fix $\bar q^{(m)}(\bar z) $ as a series expansion in $1-\bar z$, for any value of $m$, not necessarily an integer. We obtain
\begin{eqnarray}
\bar q^{(m)}(\bar z) = (1-\bar z)^{m-1} \Gamma^2(1-m)&&\left(  1+\frac{2 m^2-6 m+1}{3 m}(1-\bar z)+ \right.  \\ 
&&\left. + \frac{20 m^4-74 m^3+19 m^2+71 m-36}{90 m (m+1)}(1-\bar z)^2+\cdots \right) \nonumber
\end{eqnarray}
Logarithmic insertions can then be obtained by taking derivatives w.r.t. $m$ and setting $m$ to the desired value. We can also approach $m=0$ with a logarithmic insertion by setting $m=1$ and then applying the $\bar D_{4d}$ operator, which reduces the value by one. We obtain
\begin{equation}
\bar H^{(0,\log J)}_{2}(\bar z) =-\frac{1}{2} \frac{\log(1-\bar z)}{1-\bar z} - \frac{\gamma_e}{1-\bar z} + \log^2(1-\bar z)\left( -\frac{1}{12}+\frac{1-\bar z}{10} + \cdots \right)
\end{equation}

\subsection{A twist conformal block with global symmetry}
In the body of the paper we deal with models with $O(N)$ global symmetry. In this case the twist conformal blocks decompose in representations of this algebra. We will be interested in the following combination
\begin{equation}
\frac{1}{2} \left( H_{d-2,T}^{(m)}(u,v) - H_{d-2,A}^{(m)}(u,v) \right) \equiv H_{d-2,T-A}^{(m)}(u,v)
\end{equation}
where $T,A$ denote the symmetric traceless and anti-symmetric representations respectively. In this case
\begin{equation}
H_{d-2,T-A}^{(0)}(u,v)
=  \left(\frac{u}{v} \right)^{\frac{d-2}{2}}
\end{equation}
The construction of the sequence of functions follows exactly the same steps as before. The divergent pieces are exactly as above. On the other hand, we are interested in working out the piece proportional to $\log v \sim \log(1-\bar z)$ for $m=2$. It turns out for $m=1$ this piece is absent. For $m=2$ we find
\begin{equation}
\left. H_{d-2,T-A}^{(2)}(u,v)
\right|_{\log(1-\bar z)} = z^{\frac{d-2}{2}}\left(\frac{2 \pi  \csc \left(\frac{\pi  d}{2}\right)}{4-d} +\frac{\pi  (d-2) (d (\bar z-1)-4 \bar z+2) \csc \left(\frac{\pi  d}{2}\right)}{2 (d-4) \bar z } z+ \cdots \right)
\end{equation}
Furthermore
\begin{equation}
{\cal D}_{sat} \left. H_{d-2,T-A}^{(2)}(u,v) \right|_{\log(1-\bar z)} ={\cal C}\left. H_{d-2,T-A}^{(2)}(u,v) \right|_{\log(1-\bar z)}=0
\end{equation}
These relations will be useful in the body of the paper. 

\section{Finite support solutions}
\label{finitesupport}

In this appendix we construct all solutions to crossing symmetry for the model introduced in section \ref{genericmodel}, with finite support in the spin. We work to first order in the breaking parameter. Let us consider the decomposition in conformal blocks
\begin{equation}
{\cal G}(u,v) = 1 +\sum_{\tau,\ell} a_{\tau,\ell} u^{\tau/2} g_{\tau,\ell}(u,v)
\end{equation}
At $g=0$ the intermediate states have twist $\tau=d-2$. As we turn on $g$, these operators, together with the external operator, may acquire an anomalous dimension:
\begin{eqnarray}
\Delta_{\varphi} &=& \frac{d-2}{2} +g \gamma^{(1)}_{\varphi} +\cdots\\
\Delta_{\ell} &=& d-2 + \ell + g\gamma^{(1)}_{\ell} + \cdots,
\end{eqnarray}
The strategy to obtain a closed equation for the spectrum at order $g$ is to expand ${\cal G}(u,v)$ up to this order, and keep the terms proportional to $\log u \log v$. The same idea was used to construct solutions around generalised free fields in four dimensions in \cite{Heemskerk:2009pn}. The piece proportional to $\log u$ can only arise from the expansion of $u^{\tau/2}$:
\begin{equation}
{\cal G}^{(1)}(u,v) = \frac{1}{2} \log u \sum_{\ell=0}^L  \gamma^{(1)}_{\ell}  a_{\ell}^{(0)} u^{d/2-1} g_{d-2,\ell}(u,v) + \cdots
\end{equation}
Next, we would like to keep the piece proportional to $\log v$. A difficulty for the present case is that the explicit expression for conformal blocks is not known. We are interested in the piece proportional to $\log v$ in a small $u,v$ expansion. This piece takes the form
\begin{equation}
g_{d-2,\ell}(u,v) =  h_\ell^{(d)}(u,v) \log v+\cdots,~~~~~h_\ell^{(d)}(u,v)=\sum_{m,n=0} c^{\ell}_{m,n} u^m v^n 
\end{equation}
The coefficients $c_{m,n}$ can be computed recursively with the help of the relation 
\begin{equation}
{\cal D}_{sat} u^{d/2-1} g_{d/2-1,\ell}(u,v)=0
\end{equation}
together with the quadratic Casimir equation. Then the crossing equation takes a very simple form
\begin{equation}
\sum_{\ell=0}^L  \gamma^{(1)}_{\ell}  a_{\ell}^{(0)}  h_\ell^{(d)}(u,v) =\sum_{\ell=0}^L  \gamma^{(1)}_{\ell}  a_{\ell}^{(0)} h_\ell^{(d)}(v,u) 
\end{equation}
This is an equation for $\gamma^{(1)}_{0},\gamma^{(1)}_{2},...,\gamma^{(1)}_{L}$. We have solved this equation for different values of $L$. We find that non-trivial solutions are only possible for $d=4$. Furthermore, they take a very simple form:
\begin{equation}
\gamma^{(1)}_{0} = \alpha,~~~\gamma^{(1)}_{2}=\gamma^{(1)}_{3}=\cdots=0
\end{equation}


\end{document}